\documentclass[doublecol]{epl2} 
\usepackage[english]{babel}
\usepackage{amsmath,
amssymb,amsfonts,latexsym}
\usepackage{graphicx}
\usepackage{color}
\usepackage{subfigure}
\usepackage{afterpage}
\usepackage{booktabs}
\usepackage{blkarray}
\usepackage{cite}
\usepackage{ulem}
\usepackage{bm}
\usepackage{braket}
\usepackage{lipsum}
\usepackage{xfrac}
\usepackage{etoolbox}
\usepackage{mathrsfs} 
\usepackage{mathtools}
\usepackage[protrusion=true,expansion=true]{microtype}
\usepackage{amsmath}
\usepackage[usenames,dvipsnames,table]{xcolor}
\usepackage[colorlinks,
citecolor=blue,linkcolor=red,urlcolor=blue]{hyperref}
\usepackage{amsfonts,amsmath,amssymb}
\usepackage{soul} 
\usepackage{enumitem} 
\usepackage{times}
\definecolor{Blue}{rgb}{0.00, 0.00, 1.00}
\definecolor{Red}{rgb}{1.00, 0.00, 0.00}
\definecolor{labelkey}{cmyk}{.1,.7,0.5,0}
\newcommand{\de}{\partial}
\newcommand{\be}{\begin{equation}}
\newcommand{\ee}{\end{equation}}
\newcommand{\I}{\mathrm{i}} 

\newcommand{\qqq}{\end{document}}

\renewcommand{\arctan}{\text{arctan}}

\title{Analytical solution of a free-fermion chain \\ with time-dependent ramps}
\shorttitle{\title} 
\author{V.~Eisler$^{1,2}$ \and R.~Bonsignori$^{1,3,4}$   \and S.~Scopa$^{5}$}
\shortauthor{V.~Eisler {\it et al.}}
\institute{                    
  \inst{1}{\small Institute of Theoretical and Computational Physics, Graz University of
Technology, Petersgasse 16, A-8010 Graz, Austria}\\
\inst{2}{\small Institute of Physics, University of Graz, Universit\"atsplatz 5, A-8010 Graz, Austria}\\
\inst{3}{\small Department of Theoretical Physics, Institute of Physics, Budapest University of Technology and Economics, Műegyetem rkp. 3., H-1111 Budapest, Hungary }\\
\inst{4}{\small HUN-REN-BME Quantum Error Correcting Codes and Non-equilibrium Phases Research Group, Budapest University of Technology and Economics, Műegyetem rkp. 3., H-1111 Budapest, Hungary
}\\
\inst{5}{\small Laboratoire de Physique de l’\'Ecole Normale Superieure, CNRS, ENS \& Universit\'e PSL,
Sorbonne Universit\'e, Universit\'e Paris Cit\'e, 75005 Paris, France}}
\vspace{-1cm}
\abstract{
We provide an exact analytical solution of the single-particle Schrödinger equation for a chain of non-interacting fermions subject to a time-dependent linear potential, with its slope varied as an arbitrary function of time. The resulting dynamics exhibit self-similar behavior, with a structure reminiscent of the domain wall melting problem, albeit characterized by a nontrivial time-dependent length scale and phase. Building on this solution, we derive hydrodynamic predictions for the evolution of particle density, current, and entanglement entropy along the chain. In the special case of a sudden quench, the system develops a breathing interface region, which may be interpreted as a realization of Wannier-Stark localization, as previously suggested on the basis of hydrodynamic arguments.}

\begin{document}

\maketitle 
\section{Introduction}
Exactly solvable models in one dimension have long served as a cornerstone in the theoretical exploration of quantum many-body physics. Their role becomes particularly valuable in out-of-equilibrium settings, where even non-interacting systems can display rich and nontrivial dynamics. A paradigmatic example is that of non-interacting lattice fermions in the presence of a linear potential, where the interplay between coherent hopping and the external field gives rise to Bloch oscillations~\cite{Bloch1929} and related Wannier-Stark localization~\cite{Wannier1962}. These phenomena are well understood for static potentials and have formed the basis of several studies, including experimental tests~\cite{Holthaus1996,Geiger2018,Guo2021}.

In the equilibrium setting, progress has been made in related inhomogeneous models, such as the gradient XX chain studied in Refs.~\cite{Smith1971,MSaitoh1973,Eisler2009}, which allows for an exact characterization of local observables and entanglement. In contrast, when the linear potential acquires a time-dependent slope, the physics becomes more involved. Some special cases are known—such as quenches in the gradient XX chain~\cite{Eisler2009}, traveling impurity~\cite{Bastianello2018,Bastianello2018b}, or periodic driving~\cite{Shon1992,Pertsch1999,BarLev2024,Duffin2024,Bhakuni2018,Bhakuni2020}. However, while hydrodynamic and numerical approaches have provided valuable insights~\cite{Wendenbaum2013,Capizzi2023}, a general exact solution for the dynamics of such driven Stark-localized systems has not been fully established. {Time-dependent linear potentials can give rise to nontrivial effects that go beyond simple generalizations of the static case. Standard examples include the periodic modulation of the driving field which may lead to dynamic localization~\cite{Dunlap1986,Eckardt2009}, where the motion of a wavepacket is effectively frozen at certain parameter values of the periodic modulation, as well as related phenomena such as coherent destruction of tunneling~\cite{Holthaus1995,Holthaus1996}. These examples illustrate that driving and localization may interplay in a nontrivial way and underscore the need to explore more general driving protocols.This goal is also motivated by the growing interest in quantum control~\cite{Brif2010,Deffner2014,Glaser2015}, Floquet engineering~\cite{Bukov2015,Eckardt2017}, and transport~\cite{Bertini2021}, where the resulting nonequilibrium dynamics can have rich and unexpected behavior.}  

More generally, and even restricting to non-interacting systems, exact results in the presence of driving are remarkably scarce~\cite{Gritsev2010}. Notable exceptions include the case of a hardcore Bose gas in a time-dependent harmonic trap, see e.g. Ref.~\cite{Minguzzi2005}, which has attracted significant attention across different research fields, including condensed matter (see Ref.~\cite{Minguzzi2022} for a review) and quantum thermodynamics~\cite{Jaramillo2016}. {In this work, we provide a new exact solution for a driven quantum many-body system. }Specifically, we solve the dynamics of a tight-binding chain under an arbitrary time-dependent linear potential, giving access to the dynamics of observables such as particle density, current, and of the entanglement entropy.

\section{Model and driving protocol}
We consider a system of non-interacting fermions hopping on a one-dimensional lattice and subject to a time-dependent linear potential, with Hamiltonian
\be\label{eq:H}
\hat{H}(t) = -\frac{1}{2} \sum_{j \in \mathbb{Z}+\frac{1}{2}} \big( \hat{c}^\dagger_j \hat{c}_{j+1} + \text{h.c.} \big) + \sum_{j \in \mathbb{Z}+\frac{1}{2}} \frac{j}{\xi(t)}  \hat{c}^\dagger_j \hat{c}_j .
\ee
For convenience, sites are labeled by half-integers; $\hat{c}^\dagger_j$ and $\hat{c}_j$ denote canonical fermionic creation and annihilation operators at site $j$, satisfying $\{\hat{c}^\dagger_j, \hat{c}_k\} = \delta_{jk}$. The function $\xi(t)$ controls the time dependence of the ramp and is left arbitrary. Throughout the rest of this work, the system may be equivalently interpreted as a gas of impenetrable bosons, with fermionic Hamiltonian \eqref{eq:H} obtained through Jordan–Wigner transformation~\cite{Jordan1928}.\\

For $t<0$, the system is prepared in the ground state of \eqref{eq:H} with $\xi(t<0)=\xi_0$. For constant slope $\xi_0$, the Hamiltonian \eqref{eq:H} becomes diagonal in the eigenbasis~\cite{Smith1971,MSaitoh1973,Eisler2009}
\be\label{eq:eigen}
\Phi_k(j)=J_{j-k}(\xi_0), \quad k\in\mathbb{Z}+\frac12,
\ee
with associated eigenvalues $\omega_k=k/\xi_0$, and $J_\nu(z)$ denoting the Bessel function of first kind. The many-body ground state wavefunction is thus obtained as a Slater determinant formed by filling all single-particle modes \eqref{eq:eigen} with $k < 0$. By inspecting the ground-state two-point function~\cite{Eisler2009},
\be\label{eq:2pt-corr}
\langle\hat{c}^\dagger_i\hat{c}_j\rangle= C_{i,j}(\xi_0)=\sum_{k > 0} J_{i+k}(\xi_0) J_{j+k}(\xi_0),
\ee
one finds that the initial state exhibits nonvanishing correlations only within the interface region $|i|,|j|\leq \xi_0$, vanishing exponentially outside. In particular, the fermion density is different from $0$ and $1$ only inside the interface.\\

\begin{figure}[t]
\centering
\includegraphics[width=.8\columnwidth]{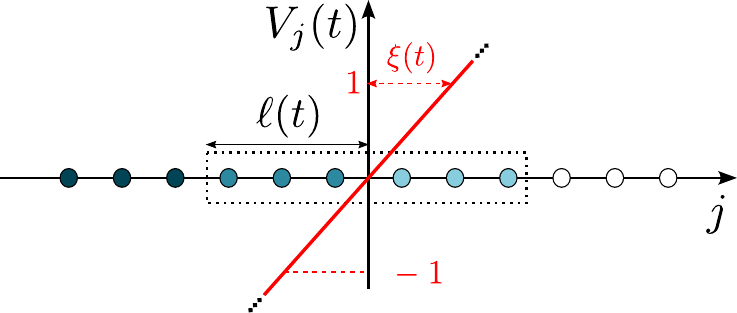}
\caption{Illustration of the linear potential $V_j(t) = j/\xi(t)$ in \eqref{eq:H}, with time-dependent slope. For $t<0$, the system features a correlated interface region $|j| \leq \xi_0$. During the driving the interface remains self-similar, with its extent given by $\ell(t)$, as obtained in the main text as a function of $\xi(t)$. Site occupations are represented using a color scale, with intensity increasing from lighter to darker shades.}\label{fig:setup}
\end{figure}

For times $t \geq 0$, the system evolves unitarily under the time-dependent Hamiltonian \eqref{eq:H}. The corresponding single-particle Schrödinger equation reads
\be\label{eq:1ptSE}\begin{split}
&\I\de_t \psi_k(j,t) = -\frac{\psi_k(j+1,t)+\psi_k(j-1,t)}{2}+\frac{j\psi_k(j,t)}{\xi(t)},
\end{split}\ee
where $\psi_k(j,t)$ denotes the time-evolved wavefunction of the $k$-th mode, $\psi_k(j,t<0)=\Phi_k(j)$. In the following, we derive an exact solution of the single-particle Schrödinger equation \eqref{eq:1ptSE}, and analyze the resulting evolution of some physical quantities under the driving. A schematic representation of the setup with the driving protocol is provided in Figure~\ref{fig:setup}.

\section{Exact solution for the modes dynamics}
Motivated by the structure of the equilibrium solution \eqref{eq:eigen}, we seek solutions of the time-dependent Schr\"odinger equation \eqref{eq:1ptSE} in the form
\be\label{eq:solution}
\psi_k(j,t) = \exp\Big(\I(j-k)\theta(t)-\I k\phi(t)\Big)  J_{j-k}[\ell(t)], 
\ee
where $\ell(t)$, $\theta(t)$, and $\phi(t)$ are functions to be determined. Inserting the ansatz \eqref{eq:solution} in eq.~\eqref{eq:1ptSE}, and using recurrence relations of Bessel functions one finds
\be\begin{split}\label{eq:tmp}
&\left[\frac{j}{\xi}-k\dot\phi + (j-k)\dot\theta-(j-k)\frac{\cos\theta}{\ell} \right]J_{j-k}(\ell)\\
&=\I\left(\dot{\ell}-\sin\theta\right)\frac{J_{j-k-1}(\ell)-J_{j-k+1}(\ell)}{2}.
\end{split}\ee
Since this equation must hold for all $j,k$, the expressions in brackets on both the left- and right-hand sides of \eqref{eq:tmp} must vanish. This yields the conditions
\be\label{eq:c1}
\dot\ell=\sin\theta; \qquad \dot\theta=\frac{\cos\theta}{\ell}-\frac{1}{\xi},
\ee
together with
\be\label{eq:dyn-phase}
\phi(t)=\int_0^t  \frac{ds}{\xi(s)}.
\ee
We note that this choice identifies $\phi(t)$ as the dynamical phase that would arise under adiabatic driving. The nontrivial aspects of the dynamics are encoded in the additional phase $\theta(t)$ and the instantaneous width of the interface $\ell(t)$, which satisfy the coupled nonlinear differential equations in eq.~\eqref{eq:c1}, with initial conditions $\ell(0)=\xi_0$ and $\theta(0)=0$.

The solution for an arbitrary driving function $\xi(t)$ can be found by introducing the variables
\be\label{eq:new-var}
u(t)=\ell(t)\cos\theta(t); \quad v(t)=\ell(t)\sin\theta(t),
\ee
such that eq.~\eqref{eq:c1} reduces to two decoupled linear ordinary differential equations
\be\label{eq:eq-for-uv}
\ddot{u}=-\frac{u}{\xi^2} -\frac{\dot\xi \dot{u}}{\xi} +\frac{1}{\xi}; \quad \ddot{v}=-\frac{v}{\xi^2} +\frac{\dot\xi (1-\dot{v})}{\xi},
\ee
with initial conditions
\be
\begin{matrix}
 u(0)=\xi_0, & \dot{u}(0)=0; \\[8pt]
 v(0)=0; & \dot{v}(0)=1-\xi_0/\xi(0).
 \end{matrix}
\ee
Note that we allow for a discontinuity $\xi_0\ne\xi(0)$, which corresponds to the quench protocol discussed later on. The differential equations admit the closed-form solutions
\begin{subequations}\label{eq:sol-uv}
\be
u(t)=\phantom{-}\xi_0\cos\phi(t) +\int_0^t ds\ \sin\big[\phi(t)-\phi(s)\big];
\ee\be
v(t)=-\xi_0\sin\phi(t) +\int_0^t ds\ \cos\big[\phi(t)-\phi(s)\big],
\ee
\end{subequations}
which depend on the driving protocol only via the dynamical phase in eq.~\eqref{eq:dyn-phase}.
In terms of these variables, the exact expression of the single-particle wavefunctions is given in eq.~\eqref{eq:solution} with $\ell=\sqrt{u^2+v^2}$ and $\theta=\arctan(v/u)$.

{We note that a related solution of the single-particle dynamics with an initially localized fermion was reported in Ref.~\cite{Dunlap1986} in the context of Stark-localized systems. However, the solution presented above is more general and provides direct access to the dynamics of the many-body problem. Indeed, using the self-similarity of the single-particle wavefunctions \eqref{eq:solution}, one can immediately write down the two-point function for the time-evolved many-body state as}
\be
\tilde C_{i,j}(t) = \langle\hat{c}^\dagger_i(t)\hat{c}_j(t)\rangle= \mathrm{e}^{\I(i-j)\theta(t)} \ C_{i,j}(\ell(t)), \label{eq:2pt-corr-t}
\ee
in terms of the ground-state correlations in eq.~\eqref{eq:2pt-corr}. The solution~\eqref{eq:2pt-corr-t} via eq.~\eqref{eq:sol-uv} thus fully characterizes the dynamics of the model \eqref{eq:H} for arbitrary driving protocols $\xi(t)$, and constitutes the main result of this work. 

\section{Adiabatic limit}
It is instructive to rewrite Eqs.~\eqref{eq:sol-uv} using the identity $\dot\phi(t) \xi(t) = 1$. After integrating by parts one obtains
\begin{subequations}\label{eq:sol-uv-2}
\be
u(t)=\xi(t) -\int_0^t ds\ \dot\xi(s) \ \cos\big[\phi(t)-\phi(s)\big];
\ee\be
v(t)=\int_0^t ds\ \dot\xi(s) \ \sin\big[\phi(t)-\phi(s)\big].
\ee
\end{subequations}
 In the quasi-static limit $\dot\xi(t) \to 0$, these expressions simplify to $u(t) \to \xi(t)$ and $v(t) \to 0$, and thus the interface follows the driving adiabatically, as expected. For slow driving protocols $\dot\xi(t)\ll 1$, the length and phase functions read
\be\label{eq:non-ad}
\ell(t)\approx u(t), \quad \theta(t)\approx \frac{v(t)}{\xi(t)}
\ee
\begin{figure}[t]
\centering
\includegraphics[width=.95\columnwidth]{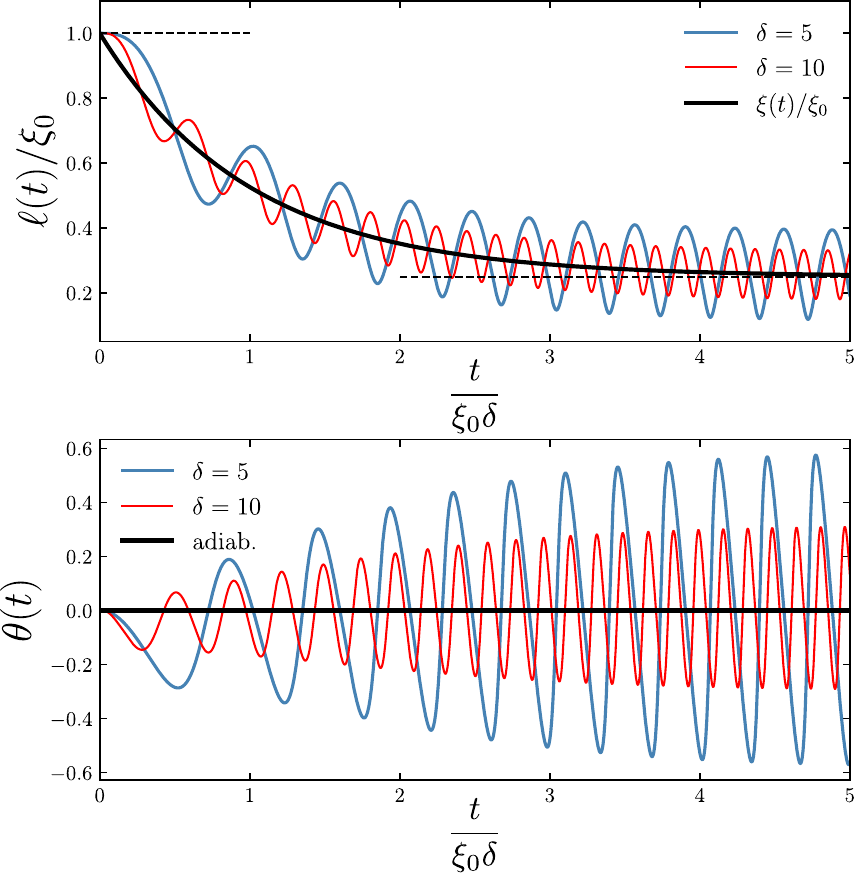}
\caption{Convergence to the adiabatic limit: Length $\ell(t)$ (top) and phase $\theta(t)$ (bottom) for the driving $\xi(t)/\xi_0 = R + (1-R) \exp\big(-\frac{t}{\xi_0\delta}\big)$, which interpolates from the initial slope $\xi_0$ to the final value $\xi_1 = R \xi_0$ ($R=0.25$ in the plots) over a timescale $\xi_0\delta$. For $\delta \gg 1$, the system approaches the adiabatic regime, with $\ell(t) \simeq \xi(t)$ and $\theta(t) \simeq 0$, up to oscillatory corrections given in eq.~\eqref{eq:non-ad}.}\label{fig:ad-lim}
\end{figure}
and feature small oscillations around the adiabatic limit. Figure~\ref{fig:ad-lim} shows the exact solution for a slow driving between the slopes $\xi_0$ and $\xi_1=R\xi_0$ (see caption) as the timescale $\delta$ of the protocol increases. The width of the interface follows the driving function, $\ell(t)\simeq \xi(t)$, with a small modulation. The phase $\theta(t)$ also shows small oscillations, with its zeros located at the extrema of $\ell(t)$, corresponding to times where the motion of the interface changes direction. 

\section{Hydrodynamic interpretation}
A coarse-grained description of the driven system emerges when the slope $\xi(t) \gg 1$. In this regime, the lattice appears effectively continuous on a large scale, allowing for a hydrodynamic treatment. The site index $j$ is then replaced by a continuous position variable $x \in \mathbb{R}$, while the mode index $k$ is replaced by the momentum in the first Brillouin zone $p \in [-\pi, \pi)$. The resulting large-scale dynamics is conveniently described by the occupation function $n_p(x,t)$, which gives the probability of finding a particle with momentum $p$ at position $x$ and time $t$~\cite{Wigner1997}. This function evolves according to the Euler hydrodynamic equation
\be
\label{eq:moyal} \partial_t n_p(x,t) + \sin p \ \partial_x n_p(x,t) = \frac{1}{\xi(t)}  \partial_p n_p(x,t),
\ee 
which is the leading-order term in the derivative expansion of the Moyal equation~\cite{Moyal1949,Fagotti2017,Fagotti2020,Essler2023}. 

As noted in Refs.~\cite{Bettelheim2012,Doyon2017}, eq.~\eqref{eq:moyal} preserves the zero entropy condition of the initial ground state, allowing the occupation function to be expressed in terms of the local Fermi points $p_\pm(x,t)$ as
\be
n_p(x,t)=\begin{cases}
1, \quad p_-(x,t)\leq p \leq p_+(x,t);\\[4pt]
0,\quad\text{otherwise}.
\end{cases}
\ee
These Fermi points $p_\pm(x,t)$ satisfy the associated equation
\be\label{eq:ghd}
\de_t p_\pm(x,t)+\sin\big[p_\pm(x,t)\big]\de_x p_\pm(x,t) =-1/\xi(t),
\ee
with initial conditions given by 
\begin{align}\label{eq:initial-cond}
&p_\pm(|x|\leq\xi_0,0) = \pm \arccos\frac{x}{\xi_0},
 \end{align}
and $p_\pm=0$ for $x>\xi_0$, $p_\pm=\pm \pi$ when $x<-\xi_0$.
\begin{figure}[t]
\centering
\includegraphics[width=\columnwidth]{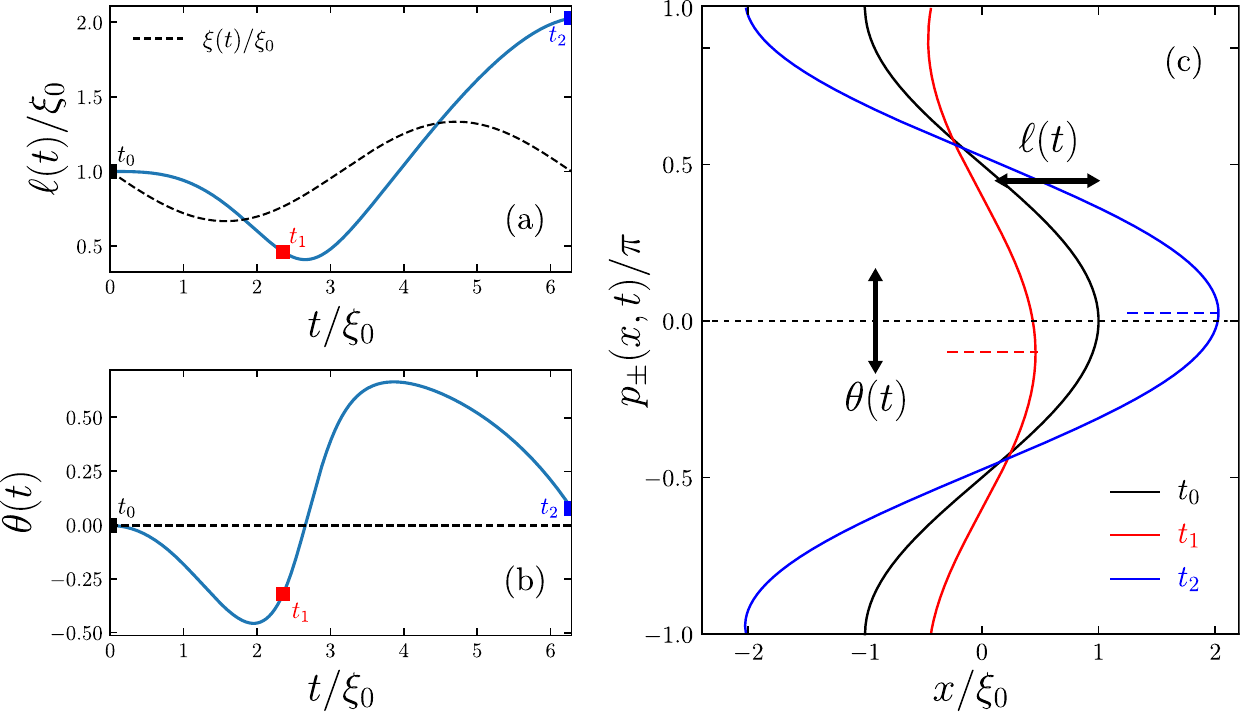}
\caption{(a)~--~Length $\ell(t)$ and (b)~--~phase $\theta(t)$ shown as functions of the rescaled time $t/\xi_0$ for the driving protocol $\xi(t) = \xi_0\bigl(1-\frac{1}{3}\sin \frac{t}{\xi_0}\bigr)$. (c)~--~Corresponding Fermi points $p_\pm(x,t)$ in \eqref{eq:fermi-pts} shown as functions of the rescaled position $x/\xi_0$ at three different times: $t_0=0$, $t_1=3\pi\xi_0/4$, and $t_2=2\pi \xi_0$ (marked with symbols in panels (a)–(b)). In the phase-space picture of panel (c), note that $\ell(t)$ controls the size of the correlated region occupied by the quasiparticles, while $\theta(t)$ determines the momentum boost imparted by the drive (marked with horizontal dashed lines).}\label{fig:fermi_pts}
\end{figure}
One can then easily check using eq.~\eqref{eq:c1} that, for times $t>0$, eq.~\eqref{eq:ghd} is solved by
 \begin{align}\label{eq:fermi-pts}
p_\pm(|x|\leq \ell(t),t) = \pm \arccos\frac{x}{\ell(t)} + \theta(t),
 \end{align} 
and $p_\pm=0$ for $x>\ell(t)$, $p_\pm=\pm \pi$ when $x<-\ell(t)$.\\

The  above result shows that the length $\ell(t)$ and phase $\theta(t)$ control, respectively, the size of the correlated region occupied by the quasiparticles and the momentum shift imparted by the drive. 
It is interesting to note that the resulting Fermi contour is related to that found for the domain wall melting (see e.g.~\cite{Antal1999,Antal2008,Karevski2002,Platini2007,Hunyadi2004,Collura2018,Gruber2019,Eisler2018,Scopa2023}), obtained by replacing $\ell(t) \to t$ and $\theta(t) \to \pi/2$. In fact, this will follow in a particular quench limit of the driving.

An example of the evolution of $p_\pm$ (and of the associated $\ell$ and $\theta$) is shown in Fig.~\ref{fig:fermi_pts}, considering a periodic modulation of the ramp's slope $\xi(t)$. The nonequilibrium character of the solution is evident from the fact that, after one period of driving, the phase-space configuration differs from its initial state.

\section{Particle density and current}
As an application, we analyze the evolution of the particle density and current, given respectively as
\be\label{eq:exact-density}
\rho(j,t)=\tilde C_{j,j}(t), \quad
{\cal J}(j,t)= \mathrm{Im}[\tilde C_{j,j+1}(t)]
\ee
via the matrix elements defined in eq.~\eqref{eq:2pt-corr-t}, where the exact expressions for $\ell(t)$ and $\theta(t)$ follow directly from the solution \eqref{eq:sol-uv}. On the other hand, the hydrodynamic limit with Fermi points given by eq.~\eqref{eq:fermi-pts} yields for the asymptotic density profile
\be\label{eq:dens}
\rho(x,t) = \int_{p_-(x,t)}^{p_+(x,t)} \frac{dp}{2\pi} = \frac{1}{\pi} \arccos \frac{x}{\ell(t)},
\ee
Similarly one finds for the current
\be\label{eq:curr}
{\cal J}(x,t) = \int_{p_-(x,t)}^{p_+(x,t)} \frac{dp}{2\pi} \sin p = \frac{\sin \theta(t)}{\pi} \sqrt{1 - \frac{x^2}{\ell(t)^2}}.
\ee
It is easy to see that Eqs.~\eqref{eq:dens}–\eqref{eq:curr} agree with the leading order approximation of the lattice result for $\ell(t) \gg 1$ \cite{Antal1999}. Furthermore, it is straightforward to verify that they satisfy the continuity equation $\partial_t \rho(x,t) + \partial_x {\cal J}(x,t) = 0$. Hydrodynamic results for higher conservation laws can be derived analogously, see e.g. Refs.~\cite{Scopa2023}.

\section{Entanglement entropy}
In addition to local observables like density and current, we study the entanglement dynamics of a single interval $A=(-\infty, j ]$ with a cut at position $j$, which probes nonlocal correlations beyond the conserved quantities discussed earlier. Entanglement can be measured by the von Neumann entropy
\be\label{eq:ent-def}
\begin{split}
S(j,t)=&-{\rm tr}\Big[ \tilde{C}_A(t) \log \tilde{C}_A(t)\\
&+\big(1-\tilde{C}_A(t)\big) \log\big(1-\tilde{C}_A(t)\big)\Big],
\end{split}
\ee
where $\tilde C_A(t)$ denotes the time-evolved two-point function \eqref{eq:2pt-corr-t} restricted to the subsystem $i,j \in A$. Since the entanglement entropy depends only on the spectrum of $\tilde{C}_A(t)$,
the phase $\theta(t)$ drops out. Consequently, the calculation of the entanglement entropy reduces to that of the domain wall melting problem, as discussed in Refs.~\cite{Dubail2017,Collura2020,Eisler2021,Scopa2021,Ares2022,Scopa2022,Scopa2022b,Scopa2023,Eisler2014}, with the only difference being the replacement of variables $t \to \ell(t)$. The final result is then
\be\label{eq:hydro-ent}
S(x,t) = \frac{1}{6} \log\left[\ell(t)\left(1 - \frac{x^2}{\ell(t)^2} \right)^{3/2} \right] + \Upsilon,
\ee
where $\Upsilon \simeq 0.4785$ is a nonuniversal constant derived from the Fisher-Hartwig conjecture in the homogeneous lattice problem~\cite{Jin2004,Calabrese2010}. \\

\begin{figure*}[t]
\centering
\includegraphics[width=\textwidth]{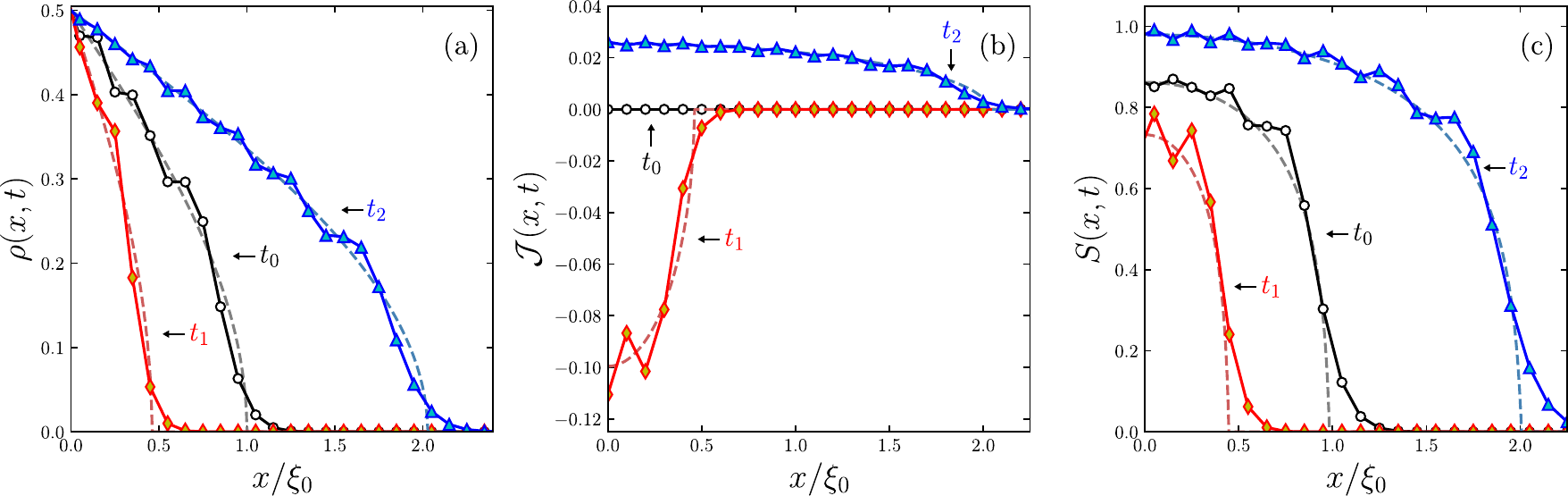}
\caption{Particle density (a), current (b), and entanglement entropy (c) under the driving protocol of Fig.~\ref{fig:fermi_pts}, shown as functions of the rescaled position $x/\xi_0$ at three times: $t_0=0$, $t_1=3\pi\xi_0/4$, and $t_2=2\pi \xi_0$. Solid lines correspond to the exact results from eqs.~\eqref{eq:exact-density} and \eqref{eq:ent-def}, while thin dashed lines show the hydrodynamic predictions from eqs.~\eqref{eq:dens}–\eqref{eq:curr} and \eqref{eq:hydro-ent}. Symbols show numerical data from exact diagonalization of the two-point correlation \eqref{eq:2pt-corr} of a finite chain with $\xi_0=10$ and $N=40$, followed by Trotter decomposition of the unitary evolution.}\label{fig:dens_curr_ent}
\end{figure*}

In Fig.~\ref{fig:dens_curr_ent} we show the particle density~(panel a), current~(b), and entanglement~(c) for the periodic driving protocol of Fig.~\ref{fig:fermi_pts}. Exact results from eqs.\eqref{eq:exact-density} and \eqref{eq:ent-def}, as well as hydrodynamic predictions from eqs.\eqref{eq:dens}-\eqref{eq:curr} and \eqref{eq:hydro-ent}, are compared with numerical data obtained via exact diagonalization of the Hamiltonian \eqref{eq:H} with slope $\xi_0$ in the single-particle basis, yielding the two-point function \eqref{eq:2pt-corr}, subsequently evolved using a Trotter decomposition of the unitary dynamics. Numerics are performed on a chain of size $N \gg \xi_0$ to avoid boundary effects. As one can see, the exact solution and numerical data are perfectly on top. The data show some visible oscillations around the hydrodynamic solutions due to the small interface length $\xi_0 = 10$, which diminish rapidly for increasing $\xi_0$.
%

\section{Quench limit}
\begin{figure}[t]
\centering
\includegraphics[width=.95\columnwidth]{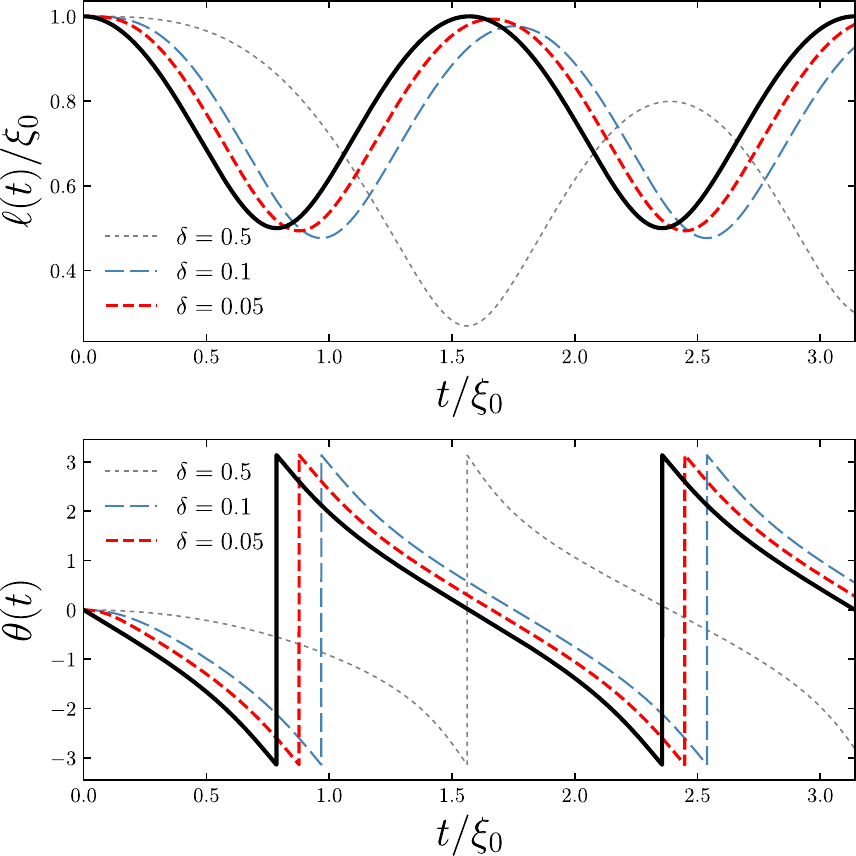}
\caption{Convergence to the quench limit.~--~Length $\ell(t)$ (top) and phase $\theta(t)$ (bottom) for the driving protocol $\xi(t)/\xi_0 = R+(1-R) \exp\big(-\frac{t}{\xi_0\delta}\big)$, $R=\xi_1/\xi_0=0.25$, also shown in Fig.~\ref{fig:ad-lim}. For $\delta \ll 1$, they converge to the results given in eqs.~\eqref{eq:l-quench}–\eqref{eq:th-quench}, shown by solid lines.}\label{fig:quench_lim}
\end{figure}

We now turn to the special case of a quantum quench, where the slope of the linear potential is suddenly changed as
\be\label{eq:quench}
\xi(t)=\begin{cases}
\xi_0, \quad t< 0;\\[4pt]
\xi_1,\quad t\geq 0.
\end{cases}
\ee
The solution to the quench problem follows as a special case of the expression given in eq.~\eqref{eq:sol-uv}. After simple algebra, one finds
\begin{subequations}\label{eq:sol-uv-quench}
\begin{align}
&u(t)=(\xi_0-\xi_1)\cos\frac{t}{\xi_1}+\xi_1;\\[4pt]
&v(t)=(\xi_1-\xi_0)\sin\frac{t}{\xi_1},
\end{align}
\end{subequations}
and thus
\be\label{eq:l-quench}
\ell(t)=\sqrt{\xi_0^2+4\xi_1(\xi_1-\xi_0)\sin^2\frac{t}{2\xi_1}};
\ee\be\label{eq:th-quench}
\tan\theta(t)=\frac{(\xi_1-\xi_0)\sin(t/\xi_1)}{\xi_1+(\xi_0-\xi_1)\cos(t/\xi_1)}.
\ee
Figure~\ref{fig:quench_lim} shows the convergence of the exact solution of the protocol in Fig.~\ref{fig:ad-lim} to the quench limit, as the driving timescale $\delta$ decreases. In this regime, the interface length $\ell(t)$ does not follow the external drive, converging instead to a periodic function with period $\tau = 2\pi \xi_1$, set by the quenched slope. The phase $\theta(t)$ also varies periodically, with apparent $2\pi$ discontinuities due to the finite Brillouin zone of the lattice. In fact, the jumps correspond to times where the interface starts to grow after shrinking, with the current changing sign.\\

Some known limits can be recovered from eqs.~\eqref{eq:l-quench}–\eqref{eq:th-quench}. In particular, $\xi_1 \to \infty$ corresponds to a protocol where the linear ramp is switched off at $t=0$, and eqs.~\eqref{eq:l-quench}-\eqref{eq:th-quench} reproduce the expressions reported in Ref.~\cite{Eisler2009}
\begin{align}
\ell(t) \overset{\xi_1\to\infty}{=} \sqrt{\xi_0^2 + t^2}, \quad \tan \theta(t) \overset{\xi_1\to\infty}{=} \frac{t}{\xi_0}.
\end{align}
By further setting $\xi_0=0$, one obtains the domain wall melting, where the formulas reduce to $\ell(t)=t$ and $\theta=\pi/2$~\cite{Antal1999,Antal2008,Karevski2002,Platini2007,Hunyadi2004,Collura2018,Gruber2019,Eisler2018,Scopa2023}. Similarly, in the limit $\xi_0 \to 0$ and finite $\xi_1$, corresponding to a domain-wall initial state quenched to a ramp of slope $\xi_1$, one recovers from eqs.~\eqref{eq:l-quench}-\eqref{eq:th-quench} the expressions
\begin{align}
\ell(t) \overset{\xi_0\to0}{=} 2\xi_1\left|\sin \frac{t}{2\xi_1}\right|, \quad \tan \theta(t) \overset{\xi_0\to0}{=} \cot\frac{t}{2\xi_1},
\end{align}
which were reported in Ref.~\cite{Capizzi2023}.\\

Hydrodynamic results in the quench limit for the particle density \eqref{eq:dens} and current \eqref{eq:curr}, and for the entanglement entropy \eqref{eq:hydro-ent}, follow directly from their general expressions upon inserting the specific form of $\ell$ and $\theta$ given in eqs.~\eqref{eq:l-quench}-\eqref{eq:th-quench}.
\begin{figure*}[t]
\centering
\includegraphics[width=\textwidth]{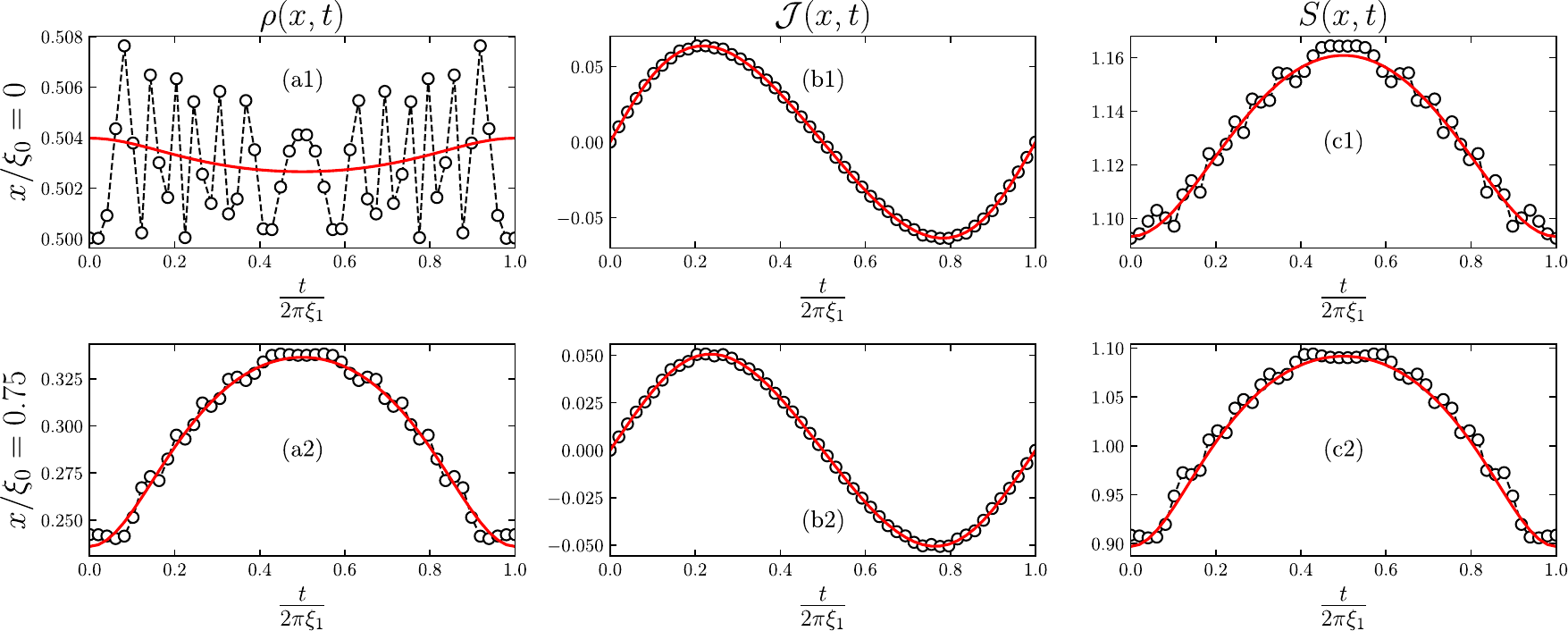}
\caption{Particle density (panels~(a1-2)), current~(b1-2), and entanglement entropy~(c1-2) during the quench $\xi_0 \to \xi_1$, $R=\xi_1/\xi_0=1.25$, shown as functions of the rescaled time $t/(2\pi\xi_1)$ at two positions: $x/\xi_0=0$ (top row) and $x/\xi_0=0.75$ (bottom row). Dashed lines correspond to the exact results from eqs.~\eqref{eq:exact-density} and \eqref{eq:ent-def}, while solid lines show the hydrodynamic predictions from eqs.~\eqref{eq:dens}–\eqref{eq:curr} and \eqref{eq:hydro-ent}. Symbols show numerical data obtained via exact diagonalization and post-quench evolution of the two-point correlation \eqref{eq:2pt-corr} for $N=160$ and $\xi_0=40$.}\label{fig:quench}
\end{figure*}
\subsection{Wannier-Stark localization}~---~It is instructive to inspect the region $\ell(t)$ hosting non-vanishing correlations during the quench. Defining the ratio $R=\xi_1/\xi_0$, the time-dependent ratio reads
\be\label{eq:l-ratio}
r(t) = \frac{\ell(t)}{\xi_0} = \sqrt{1 + 4R(R-1) \sin^2 \frac{t}{2\xi_1}}.
\ee
This resembles the result for a Tonks-Girardeau gas under a harmonic trap quench $\omega_0\to\omega_1$~\cite{Minguzzi2005,Minguzzi2022}, where
\be
r_\text{TG}(t) = \sqrt{1 + (R_\text{TG}^2 - 1)\sin^2(\omega_1 t)},
\ee
with $R_\text{TG} =\omega_0/\omega_1$ being the ratio of the trap frequencies. For the trapped gas, 
$r(t)$ oscillates between $1$ and $R_\text{TG}$, while for the chain the other extremum is given by $|2R-1|$. In particular, if $R=1/2$, i.e. the slope of the ramp is doubled, the interface compresses into a perfect domain wall at specific times, then expands back, returning to its original size with period $2\pi\xi_1$. These breathing modes in the quenched linear chain~\eqref{eq:H} reflect Wannier-Stark localization~(as discussed in \cite{Capizzi2023}) and arise because momenta are defined modulo $2\pi$ in the Brillouin zone. This feature is absent in the continuum, where coherent wavepackets drift under a constant force. The manifestation of these breathing modes is shown in Fig.~\ref{fig:quench}.

\section{Summary and conclusion}
In this work, we investigated the nonequilibrium dynamics of a tight-binding chain of non-interacting fermions (or, equivalently, of hardcore bosons) subject to a time-dependent linear potential. By obtaining an exact solution of the single-particle Schrödinger equation for arbitrary driving protocols, we analyzed some transport properties of the driven chain, and discussed their large-scale hydrodynamic description. We considered the adiabatic limit of slow driving, as well as the case of a quantum quench, corresponding to a sudden change in the slope of the linear potential, where our results generalize known findings in the literature—especially those of Refs.~\cite{Eisler2009,Capizzi2023}—and offer new insights into the physics of Wannier–Stark localization.\\

Future directions include extending this framework to interacting systems~\cite{Eisler2017}, where localization phenomena are known to persist~\cite{Schulz2019}. This feature has been recently shown even in absence of an underlying lattice structure \cite{Meinert2017}. Another promising avenue is the investigation of the entanglement Hamiltonian \cite{inprep}, following the recent developments for inhomogeneous chains~\cite{Tonni2018,Rottoli2022,Bonsignori2024,Bernard2024}. Unlike the entanglement entropy, which remains insensitive to the phase associated with the drive, the entanglement Hamiltonian is instead expected to depend non-trivially on $\theta(t)$.
\vspace{.25cm}
\subsection{Acknowledgements}~---~The authors acknowledge L. Capizzi, J. Dubail, and P. Vignolo for useful discussions. RB is supported by the HUN-REN Welcome Home and Foreign Researcher Recruitment Programme 2023. SS acknowledges support from the MSCA Grant No.~101103348 (GENESYS). This research was funded in part by the Austrian Science Fund (FWF) Grant-DOI:10.55776/P35434 and 10.55776/PAT3563424.
This work has been partially funded by the European Union. Views and opinions expressed are however those of the author(s) only and do not necessarily reflect those of the European Union or the European Commission. Neither the European Union nor the European Commission can be held responsible for them.
\bibliographystyle{eplbib}
\bibliography{bibliography.bib} 
\end{document}